\documentclass[iop]{emulateapj}

\newcommand{\kms}{{\,km\,s}$^{-1}$}
\newcommand{\teff}{{$T_\mathrm{eff}$\,}}
\newcommand{\logg}{{log~$g$\,}}
\newcommand{\vsini}{$v_{\rm eq}\sin{i}\,$}
\newcommand{\vc}{$v_\mathrm{c}\,$}
\newcommand{\vrad}{$v_\mathrm{r}\,$}
\newcommand{\vs}{$v_\mathrm{s}\,$}
\newcommand{\Msun}{\,$\rm{M}_\odot$}

\slugcomment{}

\shorttitle{Supernova disrupted binary}
\shortauthors{Dufton et al.}

\begin{document}


\title{The VLT-FLAMES Tarantula Survey: The fastest rotating O-type star and shortest period LMC pulsar -- remnants of a supernova disrupted binary?}

\author{
 P.L.~Dufton\altaffilmark{1},
 P.R.~Dunstall\altaffilmark{1},
 C.J.~Evans\altaffilmark{2},
 I.~Brott\altaffilmark{3},
 M.~Cantiello\altaffilmark{4,5},
 A.~de Koter\altaffilmark{6}
 S.E.~de~Mink\altaffilmark{7*},
 M.~Fraser\altaffilmark{1},
 V.~H\'enault-Brunet\altaffilmark{8},
 I.D.~Howarth\altaffilmark{9},
 N. Langer\altaffilmark{4},
 D.J. Lennon\altaffilmark{10},
 N.~Markova\altaffilmark{11},
 H.~Sana\altaffilmark{6},
 W.D.~Taylor\altaffilmark{8}
}

\altaffiltext{1}{Astrophysics Research Centre, School of Mathematics and Physics, Queen's University Belfast, Belfast BT7 1NN, Northern Ireland, UK}
\altaffiltext{2}{UK Astronomy Technology Centre, Royal Observatory
Edinburgh, Blackford Hill, Edinburgh, EH9 3HJ, UK}
\altaffiltext{3}{University of Vienna, Department of Astronomy,
T\"{u}rkenschanzstr. 17, A-1180 Vienna, Austria}
\altaffiltext{4}{Argelander Institut f\"{u}r Astronomie der
Universit\"{a}t Bonn, Auf dem H\"{u}gel 71, 53121 Bonn, Germany}
\altaffiltext{5}{Kavli Institute for Theoretical Physics, Kohn Hall, University of California, Santa Barbara, CA 93106}
\altaffiltext{6}{Astronomical Institute `Anton Pannekoek', University of Amsterdam, Postbus 94249, 1090 GE, Amsterdam, The Netherlands}
\altaffiltext{7}{Space Telescope Science Institute, 3700 San MartinDrive, Baltimore, MD 21218, USA}
\altaffiltext{8}{Scottish Universities Physics Alliance, Institute for Astronomy, University of Edinburgh, Royal Observatory Edinburgh, Blackford Hill,
Edinburgh, EH9 3HJ, UK}
\altaffiltext{9}{Department of Physics \& Astronomy, University College London, Gower Street, London, WC1E 6BT, UK}
\altaffiltext{10}{ESA, Space Telescope Science Institute, 3700 San MartinDrive, Baltimore, MD 21218, USA}
\altaffiltext{11}{Institute of Astronomy with NAO, Bulgarian Academy of Sciences, PO Box 136, 4700 Smoljan, Bulgaria}
\altaffiltext{*}{Hubble Fellow.}

\email{p.dufton@qub.ac.uk} 


\begin{abstract}
We present a spectroscopic analysis of an extremely rapidly rotating late O-type star, VFTS102, observed during a spectroscopic survey of 30 Doradus. VFTS102 has a projected rotational velocity larger than 500\kms\ and probably as large as  600\kms; as such it would appear to be the most rapidly rotating massive star currently identified. Its radial velocity differs by 40\kms\ from the mean for 30 Doradus, suggesting that it is a runaway. VFTS102 lies 12\,pcs from the X-ray pulsar PSR J0537-6910 in the tail of its X-ray diffuse emission. We suggest that these objects originated from a binary system with the rotational and radial velocities of VFTS102 resulting from mass transfer from the progenitor of PSR J0537-691 and the supernova explosion respectively.
\end{abstract}





\section{Introduction}
In recent years the importance of binarity in the evolution of massive stars has been increasingly recognised. This arises from most OB-type stars residing in multiple systems \citep{mas09} and the significant changes to stellar properties that binarity can cause \citep[see, for example,][]{pod92, lan08, eld11}.

Here we present a spectroscopic analysis of a rapidly rotating (\vsini $\sim 600$\kms) O-type star in the 30 Doradus region of the Large Magellanic Cloud (LMC). Designated VFTS102 \citep[][hereafter Paper~I]{eva11}\footnote{Aliases include: ST92 1-32; 2MASS J05373924-6909510}, the star is rotating more rapidly than any  observed in recent large surveys \citep{mar06, hun09} and may also be a runaway.  It lies less than one arcminute from the X-ray pulsar, PSR J0537-6910, which is moving away from it. 

We suggest that VFTS102 might originally have been part of a binary system with the progenitor of the pulsar. 

\section{Observations}

Spectroscopy of VFTS102 was obtained as part of the VLT-FLAMES Tarantula Survey, covering the 3980-5050\AA\ region at a spectral resolving power of 7000 to 8500. Spectroscopy of the H$\alpha$\ region was also available, although this was not used in the quantitative analysis. Details of the observations and initial data reduction are available in Paper~I.

The spectra were normalised to selected continuum windows using a sigma-clipping rejection algorithm to exclude cosmic rays.  No velocity shifts were observed between different epochs, although simulations \citep[see,][]{san09} indicate that 30\% of  short period (less than 10 days) and effectively all longer term binaries would not have been detected. We have therefore assumed VFTS102 to be single and the sigma-clipped merged spectrum displays a signal-to-noise ratio of approximately 130 and 60 for the 4000-4500 and 4500-5000\AA\ regions respectively.

An O9:\,Vnnne spectral classification was obtained by smoothing and rebinning the spectrum to an effective resolving power of 4000 and comparing with standards compiled for the Tarantula Survey (Sana et al. in preparation).   The principle uncertainties arise from the extremely large rotational broadening and significant nebular contamination of the \ion{He}{1} lines, with the two suffixes indicating extreme line broadening (`nnn') and an emission-line star (`e').

\begin{figure*}
\includegraphics[angle=0,scale=0.5]{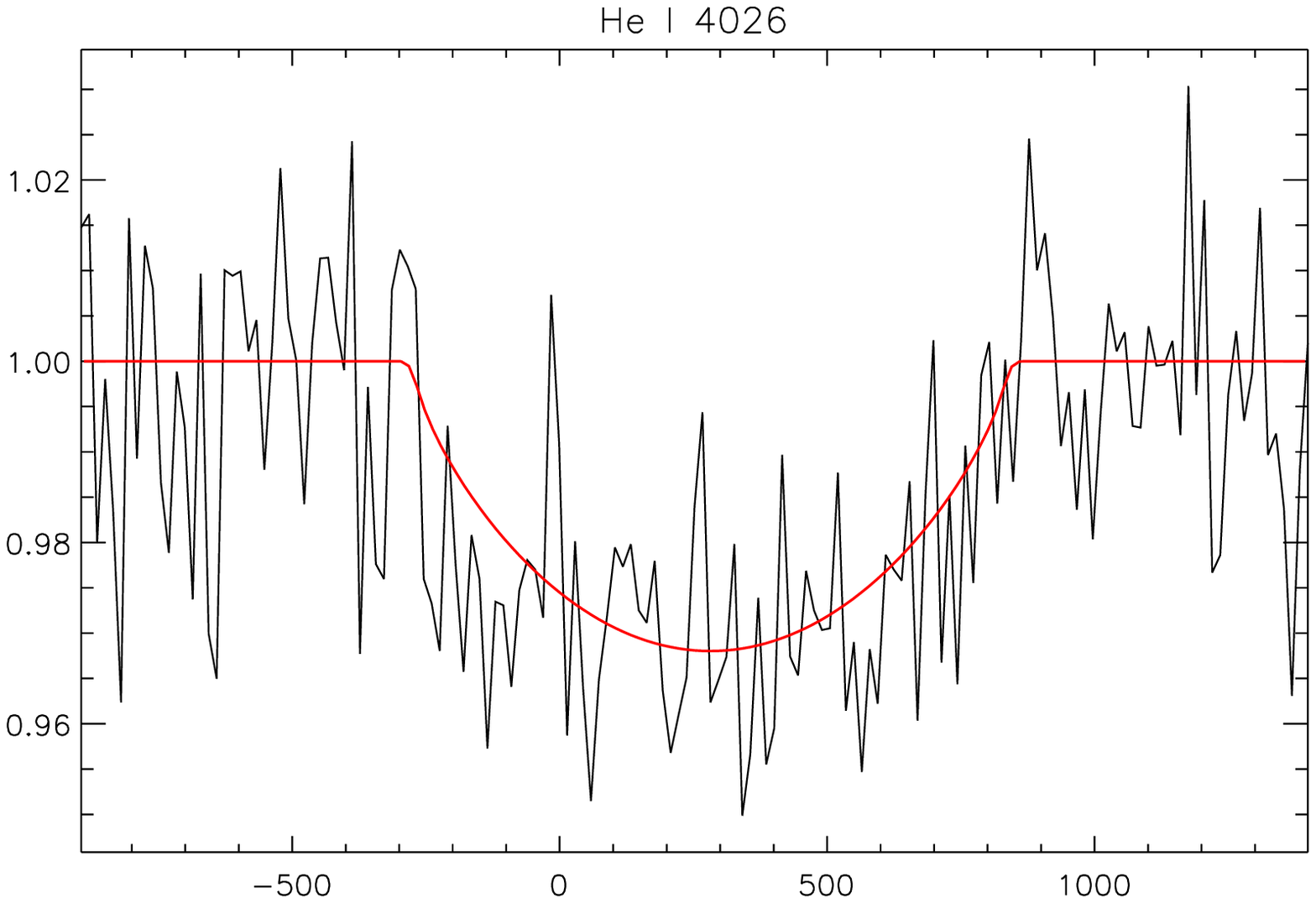}
\includegraphics[angle=0,scale=0.5]{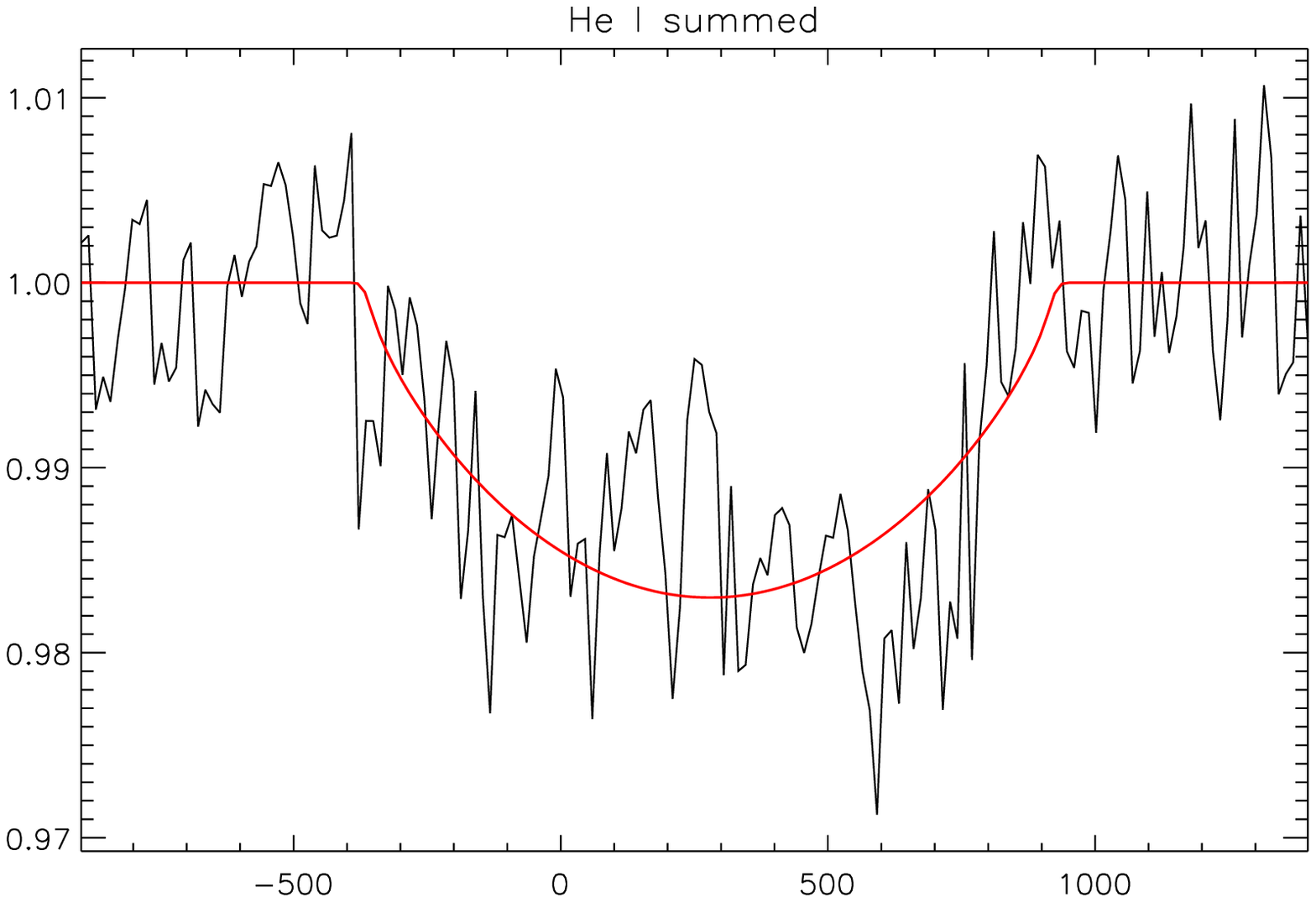}
\includegraphics[angle=0,scale=0.5]{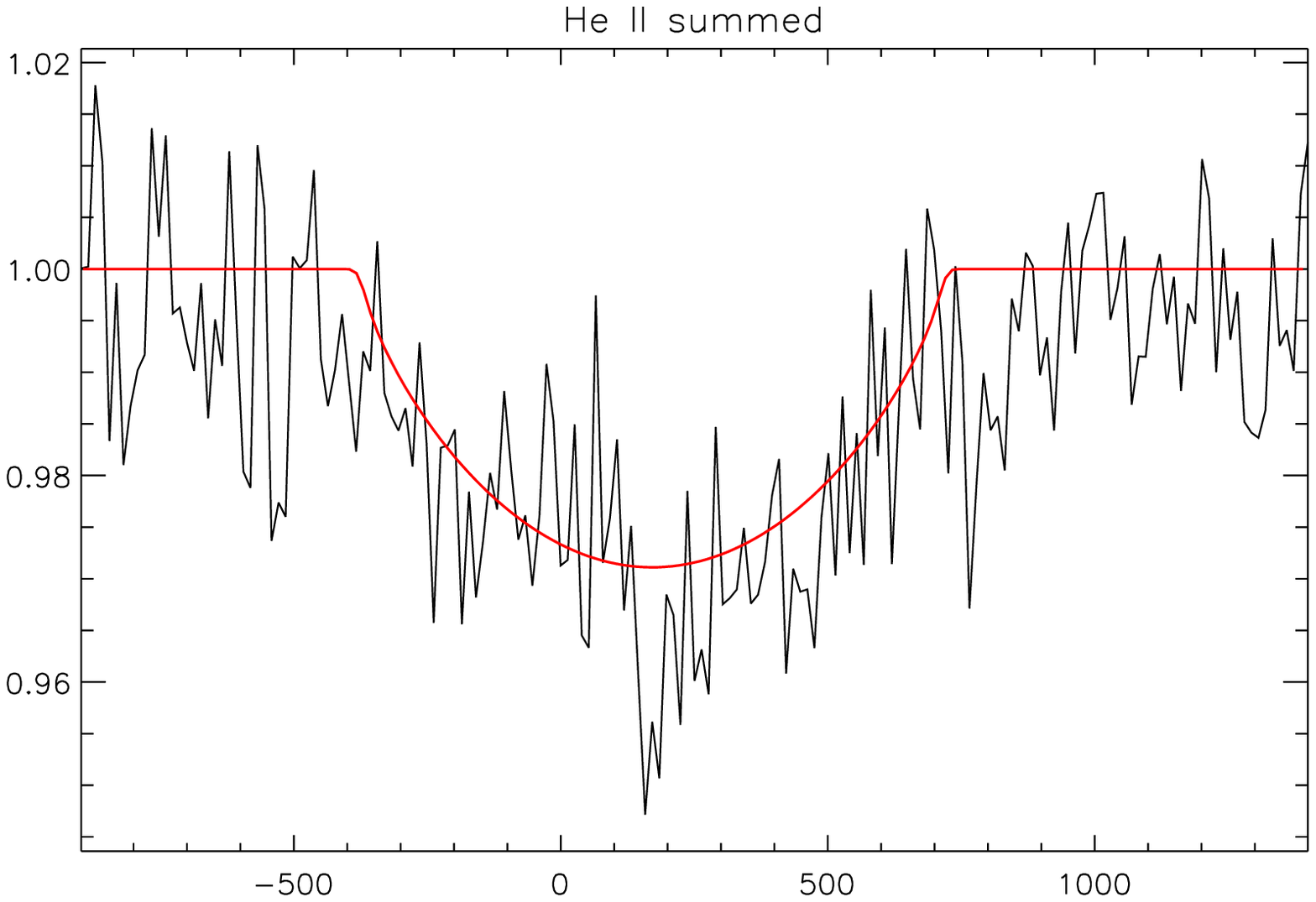}
\includegraphics[angle=0,scale=0.5]{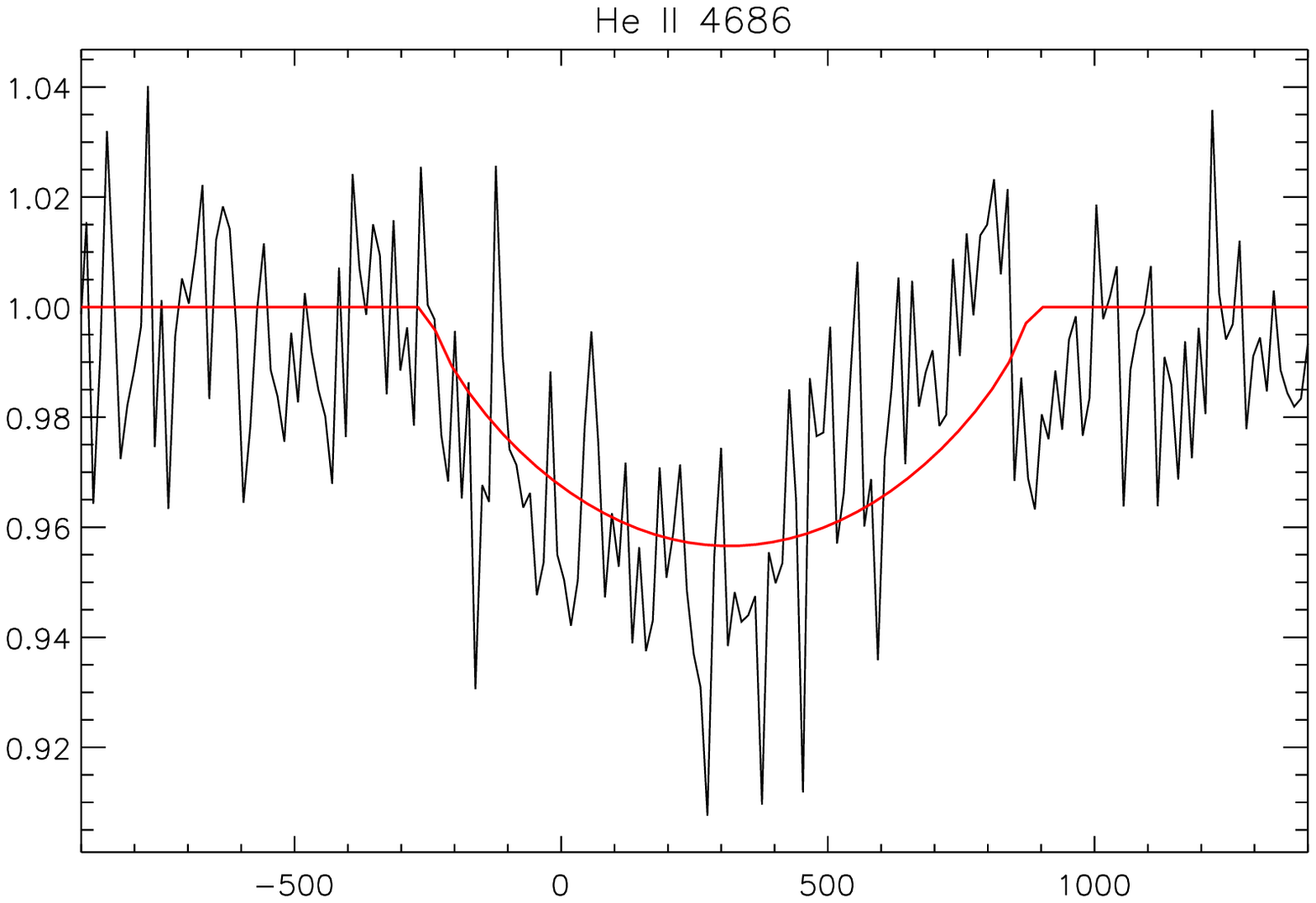}
\caption{Observed spectra (in velocity space) for VFTS102 and rotationally broadened profiles for the \ion{He}{1}  line at 4026\AA, the combined profile for the \ion{He}{1}  lines at 4026, 4143 and 4387\AA, the combined profile for the \ion{He}{2}  lines at 4200 and 4541\AA\ and the \ion{He}{2}  line at 4686\AA \label{102_1}.
}
\end{figure*}

\section{Analysis}

\subsection{Projected rotational velocity} 	\label{s_vsini}

The large rotational broadening of the spectral features makes reliable measurements of the projected rotational velocity, \vsini, difficult.  We have used  a Fourier Transform (FT) approach as discussed by \citet{sim07}, supplemented by fitting rotational broadened  profiles (PF) to the observed spectral features.

The Balmer lines have significant nebular emission and hence the weaker helium spectra were utilized, as illustrated in Fig. \ref{102_1}. The \ion{He}{1} line at 4471\AA, although well observed, also showed significant nebular emission and was not analysed. By contrast the line at 4026\AA\  showed no evidence of emission and yielded a plausible minimum in the Fourier Transform for a \vsini of 560\kms. The PF methodology leads to a slightly higher estimate (580\kms). The  \ion{He}{1} lines  at 4143 and 4387\AA\ were observed although they are relatively weak. They and the line at 4026\AA\ were converted into velocity space, merged and analysed. The two methodologies yielded effectively identical  estimates of 640\kms; a similar procedure was undertaken for the \ion{He}{2}  lines at 4200 and 4541\AA\ yielding 540\kms\ (FT) and 510\kms\ (PF). The \ion{He}{2} line at 4686\AA\  was found to be sensitive to the normalisation with a \vsini\ of $\sim$560\kms\ being estimated.

The individual results should be treated with caution but overall they imply that this star is rotating near to its critical velocity, with the mean value for the FT estimates being 580\kms. As discussed by \citet{tow04}, projected rotational velocities may be underestimated at these large velocities. For a B0 star  rotating at 95\% of the critical velocity, this underestimation will be approximately 10\%. Hence our best estimate for the projected rotational velocity is $\sim$600\kms.  A lower  limit of 500\kms\ has been adopted, whilst the upper value will be constrained by the critical velocity  of approximately 700\kms\ from the models of \citet{bro11}.

This estimate is significantly higher than those ($\la$ 370\kms) found by \citet{mar06} and \citet{hun09}  in their LMC B-type stellar samples. It is also larger than any of the preliminary estimates ($\la$450\kms) for $\sim 270$\ B-type stars in the Tarantula survey, although other rapidly rotating O-type stars have been identified. As such it would appear to have the highest projected rotational velocity estimate of any massive star yet analysed.

\subsection{Radial velocity} \label{s_vr}

Radial velocities were measured by cross-correlating spectral features against a theoretical template spectrum taken from a grid calculated using the code TLUSTY \citet{hub88} -- see \citet{duf05} for details.

Five spectral regions were considered, viz. H$\delta$\ and H$\gamma$\ (with the cores excluded); \ion{He}{1} at 4026\AA; 4630-4700\AA\ with strong multiplets due to \ion{C}{3} and \ion{O}{2} and an \ion{He}{2} line; 4000-4500\AA\ (with nebular emission being excluded).  The measurements are in excellent agreement with a mean value of 228$\pm$12\kms; if the error distribution is normally distributed the uncertainty in this mean value would be 6\kms.

From a study of $\sim$180  presumably single O-type stars in the Tarantula survey Sana et al. (in preparation) find a mean velocity of 271\kms\ with a standard deviation of 10\kms. Preliminary analysis of the B-type stars in the same survey has yielded 270$\pm$17\kms. VFTS102 lies more than two standard deviations away from these results, implying that it might be a runaway.

\begin{deluxetable}{llccccccc}
\tablecaption{Properties of VFTS102\label{t_sum}}
\tablehead{
\colhead{Parameter}       &    \colhead{Estimate} 
}
\tablewidth{0pt}
\startdata
Spectral type                                                         &  O9:~Vnnne \\
\teff (K)                                                                  &  36000$\pm$5000 \\
\logg (cm s$^{-2}$)                                                &  3.6$\pm$0.5 \\
\vsini   (km s$^{-1}$)                                                         & $600\pm 100$  \\     
\vrad (km s$^{-1}$)                                                           & 228$\pm$6       \\
log L/L$_{\odot}$                                                     & 5.0$\pm$0.2  
\enddata
\end{deluxetable}

\subsection{Atmospheric parameters} \label{s_atm}

While the equatorial regions of VFTS102 will have a lower gravity than the poles (because of centrifugal forces), and hence a lower temperature (because of von Zeipel gravity darkening),  we first characterise the spectrum by comparison with those generated with spatially homogeneous models, convolved with a simple rotational-broadening function. We have used both our TLUSTY grid and FASTWIND calculations \citep{pul05}, adopting an LMC chemical composition.  For the former, the strength of the \ion{He}{2} spectrum implies an effective temperature ($T_{\rm eff}$) of $\sim$32500--35000~K, whilst the wings of the Balmer lines lead to a surface-gravity estimate of $\sim$3.5 dex (cgs). For the latter after allowing for wind effects, the corresponding parameters are 37000~K and 3.7 dex. The helium spectra are consistent with a solar abundance but with the observational and theoretical uncertainties we cannot rule out an enhancement.

Given its projected equatorial rotation velocity, VFTS102 is almost certainly viewed at $\sin i\sim{1}$. Hence the relatively cool, low-gravity equatorial regions will  contribute significantly to the spectrum.  Although their surface flux is lower than for the brighter poles, the analyses discussed above may underestimate the global effective temperature and gravity.  However, the rotating-star models discussed below suggest that the effects are not very large.  We therefore adopt global estimates for the effective temperature of 36000~K and 3.6 dex but note that the {\em polar}  gravity could be as large as 4.0 dex.  Varying the global parameters by the error estimates listed in Table \ref{t_sum} leads to significantly poorer matches between observation and the standard models, but, given the caveats discussed above, those errors should still be treated with caution.

For  near critical rotational velocities, the stellar mass can be estimated. \citet{how01} show that the stellar mass can be written in terms of $\omega/\omega_{\rm c}$\footnote{The ratio of the equatorial angular velocity to that at which the centrifugal acceleration equals the gravitational acceleration.}, $v_{\rm eq}$ and the polar radius. Assuming that $\sin i\sim 1$ and adopting the critical velocities from our single star models, we can estimate the first two quantities. Additionally for any given value of  $\omega/\omega_{\rm c}$, the polar radius can be inferred from the absolute visual magnitude and the unreddened (B-V). The former can be estimated  from the luminosity (see Sect. \ref{absmag}) and the latter from our effective temperature estimate and the LMC broad-band intensities calculated by \citet{how11}. We find $M \gtrsim 20$~M$_\odot$\ for $v_{\rm eq} \sim 600$\kms\ and $T_{\rm eff} \la  38000$~K. Only by adopting a smaller value for $v_{\rm eq}$ can we push the mass limit down, but even with $v_{\rm eq} \sim 500$\kms\ the mass must exceed $\sim$17M$_\odot$.

\begin{figure}
\includegraphics[angle=0,scale=0.475]{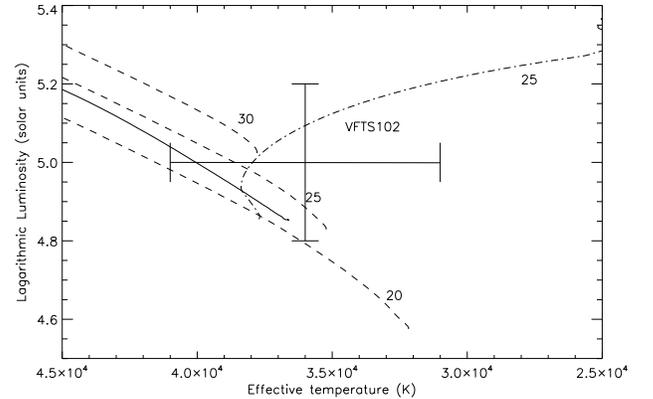}
\caption{HR-diagram showing the estimated position of VFTS102. The evolutionary tracks (identified by their mass) have rotational velocities of approximately 600\kms\ (dashed lines) and 400\kms\ (dashed-dot line). Also shown is the evolution of the secondary star following mass transfer for the binary model of \citet{can07} (solid line). 
}
\label{f_tracks}
\end{figure}

\subsection{Luminosity}\label{absmag}

From extant photometry (see Paper I), the (B-V) colour of VFTS102 is 0.35, implying an E(B-V) of 0.6 using colours calculated from our TLUSTY grid. Adopting a standard reddening law leads to a logarithmic luminosity (in solar units) of 5.0 dex, with an E(B-V) error of $\pm$0.1  corresponding to an uncertainty of  $\pm$0.1\,dex.  However there are other possible sources of error, for example deviations from a standard reddening law and hence we have adopted a larger random error estimate of $\pm 0.2$\,dex.

As VFTS102 is an Oe-type star, its intrinsic colours may be redder than predicted by our TLUSTY grid and indeed an infrared excess is found from published (de-reddened) 2MASS photometry. Inspection of a K-band VISTA image shows no evidence of contamination by nearby sources. Further evidence for circumstellar material is found in the strong H$\alpha$ emission, which is double peaked as is the nearby \ion{He}{1} line at 6678\AA, which supports our adoption of a $\sin i \sim 1$. Additionally there are weak double-peaked \ion{Fe}{2} emission features (e.g. at 4233\AA), consistent with an Oe-type classification. Unfortunately our photometry and spectroscopy are not contemporaneous but if VFTS102 was in a high state when the optical photometry was taken, we may have overestimated the luminosity of the central star \citep[see][for colour and magnitude variations of Be stars]{deW06}.

\section{Past and future evolution}

Stellar evolution calculations for both single and binary stars are available in the literature \citep[see][]{mae11}. For very fast rotation, they  suggest that rotational mixing is so efficient that stars may evolve quasi-chemically homogeneously  \citep{mae87,woo06, can07,deM09, bro11}. However, with different physical assumptions, models do not evolve chemically homogeneously even for the fastest rotation rates \citep{can07,eks08}.

\subsection{Single star evolution} \label{d_single}
Fig.~\ref{f_tracks} illustrates evolutionary tracks for  LMC single stars calculated using the methodology of \citet{bro11} for an initial equatorial rotational velocity of 600\kms, together with that for a more slowly rotating model. The former are evolving chemically homogeneously whilst the latter follows a `normal' evolutionary path. \citet{eks08} calculated models for a range of metallicities and masses between 3 and 60\Msun\ but found that the stars followed normal evolutionary paths even for near critical rotational velocities.

The estimated parameters of VFTS102 are consistent with our tracks for initial masses of $\sim$20-30\Msun. Our models show a relatively rapid increase in the surface helium abundance due to their homogeneous evolution. For example the 25\Msun\ model shows an enrichment of a factor of two after approximately 4 million years and when the effective temperature has increased to approximately 39000~K. By contrast the models of  \citet{eks08} show no significant helium abundance implying that an accurate helium abundance estimate for VFTS102 would help constrain the physical assumptions.

\subsection{Binary star evolution} \label{d_binary}

Below, we first discuss the environment of VFTS102 and then consider a possible evolutionary scenario.

\subsubsection{A pulsar near VFTS102} \label{d_pulsar}
VFTS102 lies in a complex environment near the open cluster NGC\,2060. In particular it lies close to a young X-ray pulsar PSR J0537-6910 \citep{mar98} and the Crab-like supernova remnant B0538-691 \citep{mic09}. VFTS102  has an angular separation of approximately 0.8 arcminutes from PSR J0537-6910 implying a spatial separation (in the plane of the sky) of approximately 12 pc.

\begin{figure}
\includegraphics[angle=0,scale=0.35]{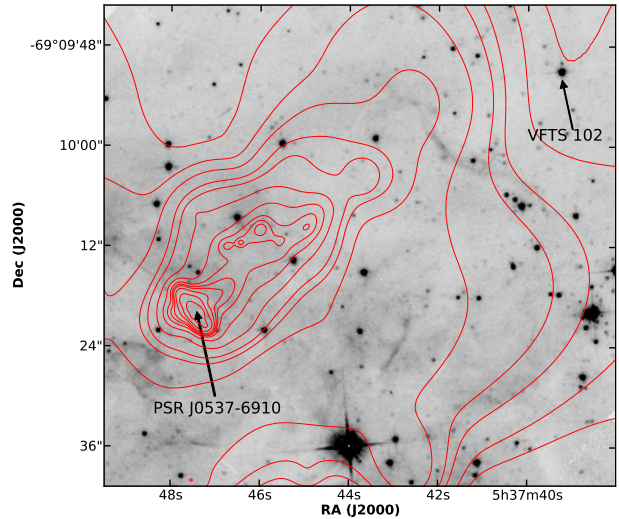}
\caption{HST WFPC2 V-band (F606W filter) image with contours from the smoothed Chandra HRC-I image overlaid. The positions of VFTS102 and PSR J0537-6910 are labeled.}
\label{f_xray}
\end{figure}

The X-ray emission consists of a pulsed localised component and a more spatially diffuse component, with the latter providing the majority of the energy. The diffuse component was identified in ROSAT  and ASCA observations by \citet{wan98a} and interpreted as coming from ram-pressure-confined material with the X-ray pulsar being identified soon afterwards by \citet{mar98}. \citet{wan98b} analysed ROSAT HRI observations and suggested that the emission could come from the remnants of a bow shock if the pulsar was moving with a 
velocity of $\sim$1000\kms. \citet{wan01} subsequently analysed higher spatial resolution CHANDRA observations, which clearly delineated this emission and implied that the pulsar 
was moving away from VFTS102. See, however, Chen et al. (2006) for an alternative explanation of the X-ray morphology, which does not imply a high velocity for the pulsar. Fig.~\ref{f_xray} superimposes these emission contours onto an HST optical image with VFTS102 being near the tail of these contours. As discussed by \citet{wan01} the spatial distribution of the diffuse X-ray emission and the SNR optical emission  are well correlated. Differences  probably arise from a foreground dark cloud and photoionization and mechanical energy input from the nearby open cluster.

Timing measurements  imply that the pulsar has a characteristic age of 5000 years \citep{mar98}, consistent with the age estimate of \citet{wan98b} from analysis of X-ray emission. 
\citet{spy02} discuss the estimation of ages from spin rates and find the results to be sensitive to both the breaking index and the composition of the pulsar core. Indeed phase connected braking index measurements for young pulsars \citep[see][and references therein]{zha01} yield breaking indices lower than the n=3 normally adopted 
with corresponding increases in the characteristic ages. Additionally, \citet{chu92} found an age of approximately 24000 years from the kinematics of the supernova remnant. 

Adopting an age of 5000 years would imply that if these objects had been part of a binary system, their relative velocity (\vs) in the plane of the sky would be approximately 2500\kms. 
Increasing this age to 24000 years would then imply \vs$\sim$ 500\kms. These values although large are consistent with a pulsar velocity of 1000\kms\ in the model of 
\citet{wan98b} and  of $\sim$600\kms\ from the separation of the diffuse X-ray and radio emission  \citep{wan01}. Additionally \citet{hob05} found a mean space velocity 
of approximately 400\kms\ for a sample of young pulsars with velocities as high as 1600\kms. From the theoretical point of view,  \citet{sto82}  found supernova kick velocities normally in excess of 300\kms, while more recently \citet{eld11} estimated kick-velocities for a single neutron star of  more than 1000\kms with a mean value of $\sim$500\kms.

\subsubsection{A binary evolution scenario for VFTS102} \label{d_scenario}

While the fast rotation of VFTS102 might be the result of the star formation process, it could also have arisen from spin-up due to mass transfer in a binary system \citep{pac81}. A subsequent supernova explosion of the donor star could then lead to an anomalous radial velocity for VFTS102 \citep{bla61,sto82}. The nearby pulsar and supernova remnant make this an attractive scenario. Of course, we cannot eliminate other possible scenarios, e.g. dynamical ejection from a cluster  \citep[see ][]{gva11} but it is unclear whether these could produce  the very large rotational velocity of VFTS102.

\citet{can07} have modelled a binary system with initial masses of 15 and 16 M$_{\odot}$ adopting SMC metallicity.  After mass transfer the primary exploded as a type Ib/c supernova. At that stage the secondary has a mass of approximately  21 M$_{\odot}$, a rotational velocity close to critical and a logarithmic luminosity of approximately 4.9 dex (see Fig.~\ref{f_tracks} for its subsequent evolution). These properties closely match the estimates for VFTS102 summarized in Table \ref{t_sum}. 

Based on grids of detailed binary evolutionary models \citep{wel01,deM07}, the initial masses of the two components of such a binary system should be comparable, with $M_2/M_1 \gtrsim 0.7$.  If the initial mass of the secondary was in the range of 14-18\Msun,   that of the primary would need to be smaller than about 25\Msun.  This agrees with the estimated initial mass of the supernova progenitor based on the kinematics of the supernova remnant \citep{mic09}. 

In this scenario, it takes the primary star about 11~Myr to evolve to the supernova stage. While the most massive stars in 30 Doradus have ages of a few million years \citep{wal99}, 
there is also evidence for different massive stellar populations with ages ranging up to about 10~Myr \citep{wal97}. Recently, \citet{dem11} have undertaken an extensive study of lower mass ($\la$4\Msun) main sequence and pre-main sequence stars in 30 Doradus. They obtain a median age of 12~Myr with ages of up 30~Myr. Hence it would appear possible that the putative binary system formed in the vicinity of 30 Doradus approximately 10~Myr ago and underwent an evolutionary history similar to that modelled by \citet{can07}.

While the evolutionary link between VFTS102 and the pulsar is an attractive scenario, the offset between VFTS102 and the apparent center of the radio emission (see Wang et al. 2001) remains to be explained. Proper motion information would be extremely valuable to further test this hypothesis. PSR J0537-6910 has not been definitely identified in other wavelength regions. \citet{mig05} 
using ACS imaging from the Hubble Space telescope found two plausible identifications that would imply an optical luminosity similar to the Crab-like pulsars. A radio survey by \citet{man06} only yielded an upper limit to its luminosity consistent with other millisecond pulsars. However estimates for both components may be obtained from the HST proper motion study (Programme: 12499; PI: D.J. Lennon) that is currently underway.

\subsection{Evolutionary future}

Irrespective of the origin of VFTS102, it is interesting to consider its likely fate. Stellar evolutionary models of rapidly rotating stars have recently been generated by \citet{woo06} and \citet{yoo06}. The latter  consider the fate of objects with rotational velocities up to the critical value (\vc). The evolution is shown to depend not only on initial mass and rotational velocity but also on the metallicity. In particular GRBs are predicted to occur only at sub-solar metallicities.

Based on our single star models, VFTS102 has a rotational velocity above $\sim 0.8$\vc\ and is thus expected to evolve quasi-chemically homogeneously. While \citet{yoo06} and \citet{woo06} estimate the metallicity threshold for GRB formation from chemically homogeneous evolution to be somewhat below the LMC metallicity, the latter note its sensitivity to the mass loss rate \citep{vin05}. Indeed all our most rapidly rotating $20-30\,M_{\odot}$\ models  are evolving chemically homogeneously throughout core hydrogen burning (Fig.~\ref{f_tracks}), a prerequisite to qualify for a GRB progenitor. In any case, within the context of homogeneous evolution VFTS102 is expected to form a rapidly rotating black hole, and a Type~Ic hypernova. This conjecture remains the same within the binary scenario of \citet{can07}. 

Assuming a space velocity of 40\kms\ for VFTS102 (compatible with its anomalous radial velocity), our evolutionary models imply that VFTS102 will travel $\sim$300-400 pc before ending its life. This is consistent with the finding of \citet{ham06} that the locations of three nearby GRBs were found several hundred parsecs away from their most likely progenitor birth locations \citep[see, however,][]{mar07,wie07,han10}.

\section{Conclusions}

VFTS102 has a projected rotational velocity far higher than those found  in previous surveys of massive stars in the LMC, and indeed it would appear to qualify as the most rapidly rotating massive star yet identified. With a luminosity of $10^5$\,L$_{\odot}$ we estimate its current mass to be approximately 25\Msun. Its extreme rotation, peculiar radial velocity,  proximity to the X-ray pulsar PSR~J0537-6910 and to a supernova remnant suggest that the star is the result of binary interaction. 

It is proposed that VFTS102 and the pulsar originated in a binary system with mass transfer spinning-up VFTS102 and the supernova explosion imparting radial velocity kicks 
to both components. If evolving chemically homogeneously, as suggested by recent models, VFTS102 could become a GRB or hypernova at the end of its life. Additionally it may provide a critical test case for chemically homogeneous evolution.

\acknowledgements

SdM acknowledges NASA Hubble Fellowship grant HST-HF- 51270.01-A awarded by STScI, operated by AURA for NASA, contract NAS 5-26555. NM acknowledges support from the Bulgarian NSF (DO 02-85). We would like to thank Eveline Helder, Paul Quinn, Stephen Smartt, Jorick Vink and Nolan Walborn for useful discussions. This paper makes use of spectra obtained as part of the VLT-FLAMES Tarantula Survey (ESO programme 182.D-0222).

{\it Facilities} \facility{VLT:Kueyen (FLAMES)}

\clearpage

\end{document}